\documentclass[aps, amsfonts, nofootinbib,notitlepage, preprintnumbers]{revtex4-1}
\usepackage{hyperref}
\hypersetup{
colorlinks=true,
linkcolor=red,
citecolor=blue,
urlcolor=blue}
\usepackage{epsf,epsfig}
\usepackage{amscd}
\usepackage{amsmath}
\input xy
\xyoption{all}

\newcommand{\beq}{\begin{equation}}
\newcommand{\eeq}{\end{equation}}
\newcommand{\bea}{\begin{eqnarray}}
\newcommand{\eea}{\end{eqnarray}}

\newcommand{\nn}{\nonumber}
\newcommand{\benn}{\begin{displaymath}}
\newcommand{\eenn}{\end{displaymath}}

\newcommand{\Dv}{{\mathbf{D}}}
\newcommand{\Ab}{{\mathbf{A}}}
\newcommand{\pb}{{\mathbf{p}}}

\def\slashchar#1{\ensuremath{                               %
   \setbox0=\hbox{${}#1{}$}       
   \dimen0=\wd0                                 
   \setbox1=\hbox{/} \dimen1=\wd1               
   \ifdim\dimen0>\dimen1                        
      \rlap{\hbox to \dimen0{\hfil/\hfil}}      
      {}#1{}                                    
   \else                                        
      \rlap{\hbox to \dimen1{\hfil${}#1{}$\hfil}}   
      /                                         
   \fi}}                                        %

\begin{document}

\title{Nuclear condensate and helium white dwarfs} 
\author{Paulo F.~Bedaque\footnote{{\tt bedaque@umd.edu}}} 
\author{Evan Berkowitz\footnote{{\tt evanb@umd.edu}}}
\affiliation{Maryland Center for Fundamental Physics,\\ Department of Physics,\\ University of Maryland, College Park, MD USA} 
\author{Aleksey Cherman\footnote{{\tt a.cherman@damtp.cam.ac.uk}}}
\affiliation{Department of Applied Mathematics and Theoretical Physics,\\ University of Cambridge, Cambridge CB3 0WA, UK }

\preprint{UM-DOE/ER/40762-507}
\preprint{DAMTP-2011-90} 

\begin{abstract}
We consider a high density region of the helium phase diagram, where the nuclei form a Bose-Einstein condensate rather than a classical plasma or a crystal.  Helium in this phase may be present in helium-core white dwarfs. We show that in this regime there is a  new gapless quasiparticle not previously noticed, arising when the constraints imposed by gauge symmetry are taken into account.  The contribution of this quasiparticle to the specific heat of a white dwarf core turns out to be comparable in a range of temperatures to the contribution from the particle-hole excitations of the degenerate electrons.   The specific heat in the condensed phase is two orders of magnitude smaller than in the uncondensed plasma phase, which is the ground state at higher temperatures, and four orders of magnitude smaller than the specific heat that an ion lattice would provide, if formed.  Since the specific heat of the core is an important input for setting the rate of cooling of a white dwarf star, it may turn out that such a change in the thermal properties of the cores of helium white dwarfs has observable implications.
\end{abstract}
\maketitle

\section{Introduction}
The behavior of matter at  high pressure is extremely rich, featuring the interplay of electromagnetic, statistical, quantum, and relativistic physics \cite{Drake:PhysicsToday}.  Aside from the intrinsic theoretical appeal of such a playground, the properties of matter in extreme conditions are important in many phenomenological applications.  Perhaps foremost among these are applications in astrophysics and planetary science, as well as in terrestrial experiments on high-energy-density matter, for instance using inertial confinement or diamond anvil cells.   Here we will discuss the behavior of helium under pressures high enough such that the average interparticle spacing $l$ is smaller than the Bohr radius $a_{0} \sim 10^{5}\, \mathrm{fm}$.  At such densities the helium gas becomes completely ionized, so that the system can be described as a plasma of helium nuclei and electrons interacting via electromagnetism.   Such conditions are expected in helium-core white dwarfs.  Helium white dwarfs (He WDs) are thought to be formed when a large part of a red giant star's envelope is removed prior to helium ignition, exposing the helium core\cite{1975MNRAS.171..555W,1991ApJ...381..449D,1993ApJ...407..649C}.   A fair number of He WDs have recently been discovered\cite{2001ApJ...553L.169T,2009ApJ...699...40S}.  A particularly interesting aspect of these observations is the discovery that the sequence of the He WDs found in \cite{2009ApJ...699...40S} comes to an early end compared to the sequence of carbon-oxygen white dwarfs.   It may  be that the explanation for this phenomenon involves developing a better understanding of the envelopes of the He WDs\cite{2009ApJ...699...40S,2003ApJ...586.1364H}, but it is also interesting to examine other alternatives, such as changes in the cooling rates of the WDs at lower temperatures due to phase transitions in the core.   Hence our goal in this paper is to develop a better understanding of the ground state of the helium plasma in the cores of He WDs.

Whether a plasma of nuclei and electrons is the ground state of a system depends on the density and temperature.  There are two relevant possibilities for the behavior of high-density helium at lower temperatures if it cannot remain a standard plasma:  crystallization or Bose-Einstein condensation of the helium nuclei.  Which one is preferred depends on the density and temperature.  For heavier white dwarfs, crystallization is the accepted model for low enough temperatures.  The standard model for He WD evolution is based on assuming that the plasma of helium nuclei and electrons survives to low temperatures, see e.g. \cite{1997ApJ...477..313A}.  However, it was recently pointed out that in He WDs it is quite plausible that the conditions are such that the helium ions Bose-condense, with potentially observable consequences for the cooling rates of the stars\cite{Gabadadze:2008mx,Gabadadze:2009dz,Gabadadze:2009jb}. (For some early work on ion condensation in general see \cite{0305-4470-36-22-341,Ashcroft:kx,PhysRevLett.95.105301,Oliva:1984fk,Oliva:1984uq,PhysRevLett.66.2915,1998MPLA...13..987L}.)  In this paper we will examine the consequences of helium ion condensation on the thermodynamic properties of the system, by developing an understanding of the spectrum of low-energy quasi-particle modes in the condensed phase.  

Let us now review the arguments that suggest that a helium plasma may Bose-condense at high enough densities.   If the Coulomb energy $E_{c}$ of the ions is much larger than their thermal energy $E_{T}~$ and quantum effects can be neglected\footnote{\label{footnote:quantum}Quantum fluctuations may become important at high enough densities for very small temperatures, and can be expected to melt the lattice when the quantum zero-point energy of the ions $E_{Q}$ overwhelms the Coulomb energy keeping them localized.  The density at which this happens is sensitive to the value that $E_{C}/E_{Q}$ must reach to trigger this quantum phase transition, and requires detailed calculation, see e.g. \cite{2004Natur.431..669B} in the case of hydrogen.  Ref.~\cite{Gabadadze:2008mx} gives some plausible arguments that quantum effects melt the crystal at $T=0$ in He WDs, but fortunately for our purposes here it will not be important to take a position on the $T=0$ ground state in He WDs.}, the ions should be expected to crystallize.  To get an estimate of the crystal-melting temperature, one can take $E_{c}$ to be dominated by nearest-neighbor interactions, so that $E_{c} = Z^{2}\alpha/l$, where $Z$ is the ion charge, $\alpha \approx 1/137$ is the fine structure constant, and $l$ is related to the density $n$ through $3/(4\pi l^{3})  = n$, while the thermal energy in the plasma phase is dominated by the contribution of the ions and can be approximated by that of a free bose gas $E_{T} \approx  T$.  (We work with natural units where $k_{B} = c=\hbar = 1$ except where stated otherwise.)  The crystal-melting temperature $T_{\textrm{melt}}$ can then be estimated from the phenomenological relation $E_{c}/E_{T} \sim 180$ \cite{1975ApJ...200..306L,PhysRevA.21.2087,1993ApJ...414..695C} , giving $T_{\textrm{melt}} \sim (a_{0}/l) 7000\,\mathrm{K}$.   

On the other hand, if $l$ becomes comparable to the thermal deBroglie wavelength $l_{DB} = \sqrt{2\pi/ MT}$, where  $M$ is the ion mass, one should expect the ions to Bose-condense.  The critical temperature $T_{c}$ of such a nuclear Bose-Einstein condensate can be estimated 
by comparing this thermal wavelength to the inter-ion distance. This woud suggest a critical temperature of $T_c\approx 2\pi/(M l^2)\approx 6.3/(Ml^2)$. A free gas would condense at a smaller temperature, $T_c = [4\pi \zeta(3/2)/3]^{-2/3} \times 2\pi/(Ml^2)\approx 1.27 /(Ml^2)$. Repulsive interactions, however,
will increase the critical temperature\cite{Huang:1999zz}. Some estimates\cite{Gabadadze:2009jb} suggest $T_c\approx 4\pi^2/(3 Ml^2)\approx 13.2/(Ml^2)$, a result qualitatively supported by more detailed calculations\cite{Rosen:2010es}.   In any case, at a certain critical density, $T_{c}$ will exceed $T_{\textrm{melt}}$, and a Bose-condensed region of the phase diagram will open up.

\begin{figure}[tbp]
\centerline{ {\epsfxsize=3.0in\epsfbox{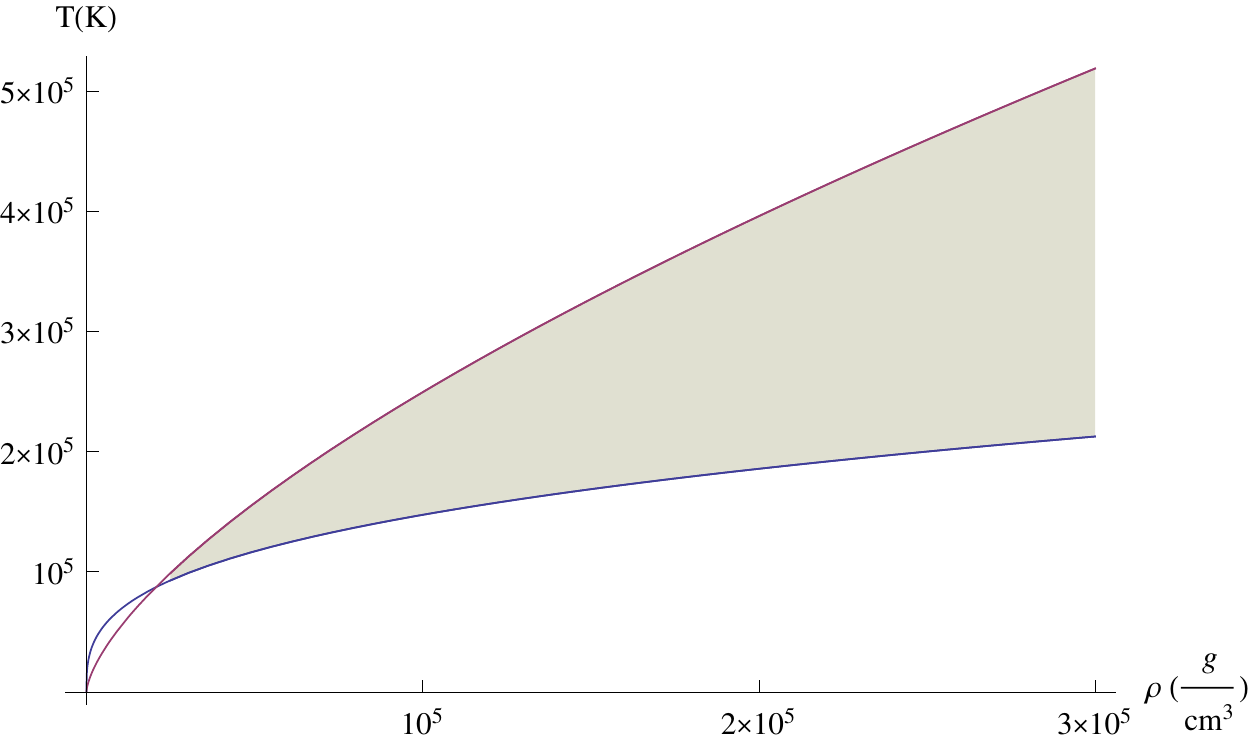}} }
 \noindent
\caption{(Color online.) Estimated crystallization (blue) and condensation (red) temperatures as a function of the WD core density, assuming the estimate for $T_{c}$ in ~\cite{Gabadadze:2009jb}.  The crystallization temperature estimate in the plot assumes quantum effects are negligible, as discussed in Footnote~\ref{footnote:quantum}.   Bose-Einstein condensation of the nuclei is expected in the shaded area. }
\label{fig:phasediagram}
\end{figure}  

Numerically, assuming the estimate in~\cite{Gabadadze:2009jb} of $T_{c} \approx 13.2/(Ml^{2})$,  the critical value of $(a_{0}/l)$ where the condensation region opens up is given by $(a_{0}/l)\approx 12.4 $, while using $T_{c}\approx 6.3/(Ml^{2})$ gives the intersection value $a_{0}/l \approx 26$.  This translates to a critical mass density of around $2 \times 10^{4}\,\mathrm{g/cm^{3}}$ to $2\times 10^{5}g/cm^{3}$, respectively.   Meanwhile,  helium white dwarf models show that their central mass density varies between $1\times10^{5}\,\mathrm{g/cm^{3}}$ to $1 \times 10^{6}\,\textrm{g/cm}^{3}$, see e.g. \cite{2000A&A...353..970P,Benvenuto:2011fj}.  These estimates make it quite plausible that the conditions appropriate for Bose-condensation of helium nuclei may be present in the cores of Helium white dwarfs at some point during their thermal evolution.  Despite (or perhaps because of) the uncertainties in the estimates, there is enough motivation to ask what the consequences of nuclear condensation may be for the bulk properties of a helium white dwarf, so that one can eventually decide whether the effects of nuclear condensation in He WDs may be detectable in current or future observations.    The region of the phase diagram where Bose-Einstein condensation is likely to occur is shown in Fig.~(\ref{fig:phasediagram}).

At low temperatures, the thermodynamic properties of a system are determined by the spectrum of low-energy excitations.  Any gapless or nearly-gapless quasiparticle modes make particularly important contributions to the thermodynamic properties, since the contributions of gapped quasiparticle modes become Boltzmann-suppressed once the temperature is low enough.  Hence the first essential step in characterizing the physics of a quantum liquid is to understand the spectrum of quasiparticle excitations, and this is the focus of this paper. The most phenomenologically important result of our analysis is that the nuclear condensate has a previously undiscovered gapless quasiparticle mode.  To find this mode it is crucial to pay careful attention to the constraints imposed by gauge invariance, and to characterize its properties correctly it is necessary to go beyond the Thomas-Fermi approximation in treating the electrons.  Armed with the full quasiparticle spectrum, which turns out to be rather different in its details than appreciated in previous studies\cite{Gabadadze:2008mx,Gabadadze:2009jb,Gabadadze:2009dz,Mosquera2010119}, we compute the specific heat of the helim-ion quantum liquid.  We find that the specific heat of this quantum liquid is very small, far smaller than the specific heat of an ion crystal or a classical gas of ions.  Fortunately, this means that our results do not qualitatively conflict with the promising phenomenological arguments regarding helium white dwarf cooling of Refs.~\cite{Gabadadze:2008mx,Gabadadze:2009jb,Gabadadze:2009dz}.  Thus if the cores of helium-rich white dwarfs undergo a phase transition to a phase with a nuclear condensate, one should expect a large drop in the specific heat of the cores, with a consequent increase in the rate of cooling of the stars.  If this increase in the cooling rate is rapid enough, nuclear condensation may help explain the early termination of the He WD sequence observed by \cite{2009ApJ...699...40S}, but whether this is indeed the case is not yet clear as we discuss in the conclusions.

The rest of this paper is organized as follows.  Section~\ref{sec:EFT} contains a description of the effective field theory describing the condensed phase, and presents a perturbative computation of the quasiparticle spectrum.  Section~\ref{sec:Corrections} discusses the conditions under which corrections to the spectrum will be small, and is followed by a discussion of the implications of the quasiparticle spectrum for the specific heat of the condensed phase in Section~\ref{sec:SpecificHeat}.  Finally, Section~\ref{sec:Conclusions} contains the conclusions and a sketch of some directions for future work.

\section{Effective Theory and Quasiparticles}
\label{sec:EFT}
Our main physics goal in this paper is to compute the specific heat of the nuclear condensate.  The general principles of effective field theory and thermodynamics imply that the specific heat of a quantum liquid can be calculated from the dispersion relations of the low-energy quasiparticle excitations of the liquid.  If the quantum liquid is the $T=0$ ground state, then the dispersion relations of quasi-particle excitations will be perturbatively close to the $T=0$ spectrum at low temperatures, so that in calculating the specific heat at small $T$ one could use dispersion relations calculated at $T=0$. In our case, however, the ground state of the system at $T=0$ may be a crystal, not a quantum liquid.   In such a case, to compute the specific heat for $T_{\textrm{melt}}< T \ll T_{c}$, one should use the quasi-particle spectrum computed at $T=0$ with the (in this case) artificial assumption that a condensate exists at $T=0$.  Temperature effects on the quasiparticle dispersion relations appear, together with other corrections, at loop level, and will be neglected here. Because the transition between the crystal and liquid phase (if there is one) must be first-order, there is no risk of finding unstable modes by perturbing around the condensed phase.  Thus we will carry out our analysis at $T=0$ and assume that the helium nuclei are in a Bose-Einstein condensate in order to compute the quasi-particle spectrum in the condensed phase and learn about its thermodynamic properties for $T\ll T_{c}$. 
 
At the densities and temperatures of interest the typical interparticle distance $l$ lies in the range $R_N \ll l \ll a_0$ where $R_N\approx 2\,\mathrm{fm}$ is the size of the $\alpha$-particle, $l$ is related to the density of ions $n$ by  $4\pi l^3 n/3=1$, and $a_0=(\alpha m)^{-1}$ with $m$ the mass of the electron is the Bohr radius. At these densities the helium nuclei can be treated as a point particles. An effective theory treating the ions, electrons and photons is described by the Lagrangian in the second quantized form
\begin{align}
\label{eq:lag_initial}
\mathcal{L} = \psi^\dagger \left( iD_0+ \mu + \frac{\Dv^2}{2M}\right)\psi - \frac{1}{4}F_{\mu\nu}F^{\mu\nu} +\bar\chi\left( iD_\mu \gamma^\mu + \mu_e \gamma^0+m\right) \chi+\cdots,
\end{align}
with $D_\mu\psi = (\partial_\mu - i Ze A_\mu)\psi$, $D_\mu\chi= (\partial_\mu + i e A_\mu)\chi$ and $Z=2$. The fields for the helium nuclei and electrons are $\psi$ and $\chi$ respectively, and $\mu$ and $\mu_e$ their chemical potentials. Omitted from eq.~(\ref{eq:lag_initial}) are terms describing the strong force interaction between nuclei since they have a very short range, and Coulomb repulsion makes them irrelevant at the temperatures and densities we will be considering.  

To study the quasiparticle spectrum, we find it convenient to integrate out the electrons. The result is that the explicit fermion term above disappears and is replaced by a trace-log term coming from the fermion determinant:
\bea\label{eq:Selectron}
\mathcal{S}_{\mathrm{electron}} &=& {\rm tr}\log\left( iD_\mu \gamma^\mu + \mu_e \gamma^0+m \right)\nn\\
&=&  {\rm tr}\log\left( i\partial_\mu \gamma^\mu + \mu_e \gamma^0+m \right)-Z e n A_{0}
+\frac{1}{2} \int\frac{d^4p}{(2\pi)^4}A^\mu(-p)  \Pi_{\mu\nu}(p)A^\nu(p) + \mathcal{O}(A^3),
\eea where $Zn$ is the electron density in an electrically neutral system. In the second line above we expanded the determinant in powers of the photon field.  The first term above is of order $\mathcal{O}(A_{\mu})^{0}$ and gives the free energy of a free electron gas. It makes an important  contribution to the thermodynamics of the system and is associated with some gapless quasiparticles, the particle-hole excitations.  However, this free contribution will not play a direct role in our discussion of the remaining excitations. The second term is the coupling between the electric potential and the electron charge that will cancel against the electrostatic energy of the ions. The third term above is of order $\mathcal{O}(A_{\mu})^{2}$, and takes the form of a photon polarization tensor whose explicit form will be given below. Terms with higher powers of the photon field will not be needed in our discussion. The polarization tensor satisfies some relations as a consequence of gauge invariance. In particular,  charge conservation implies that $p^\mu \Pi_{\mu\nu}(p)=0$. This relation, combined with the requirements of rotational invariance (but not boost invariance, which is broken by the electron chemical potential term), implies that $\Pi_{\mu\nu}$ is given by two scalar functions of $p_0$ and $\pb^2$, $\Pi(p_0, \pb^2)$ and  $\Pi^{T}(p_0, \pb^2)$ as (see appendix)
\beq
\Pi_{\mu\nu} =
\begin{pmatrix}
\Pi  &   -\frac{p_ip_0}{\mathbf{p}^2}\Pi \\
  -\frac{p_j p_0}{\mathbf{p}^2}\Pi  &      \frac{p_i p_j p_0^2}{\mathbf{p}^4}\Pi +
  (p_i p_j-\delta_{ij}\mathbf{p}^2)\Pi^T
\end{pmatrix}
\eeq

We will assume that the dynamics does not prevent the condensation of the bosonic ion field $\psi$, in accord to the general discussion above, and parametrize  $\psi$ as $\psi=(v+h) e^{i\phi}$. In order not to confuse  excitations of physical degrees of freedom with the illusory excitations of gauge-dependent quantities we need to fix the gauge. We use the Fadeev-Popov method, choosing a variation of the $R_\xi$-gauge. The gauge-fixing term to be added to the Lagrangian is
\beq
\mathcal{L}_{gauge} = -\frac{1}{2\xi}\left(
\nabla.\Ab - \frac{2M}{ev}\partial_0 h - \frac{\xi Zev^2}{M} \phi 
\right)^2
\eeq 
This gauge-fixing term cancels the quadratic terms mixing $\phi$ with $h$ or $\Ab$. With our choice of field parametrization and gauge-fixing term, the Fadeev-Popov ghost fields are decoupled from $A_{\mu}, h, \phi$, and we can neglect them for our purpose of computing the spectrum.   Up to quadratic order in the fields  the Lagrangian becomes
\begin{align}
\mathcal{L}_{\mathrm{quad}} &=2\mu v h +Ze(v^2-n)A_0 - \frac{1}{2M}(\nabla h)^2+\mu h^2+2 Ze v A_0 h -\frac{2M^2}{\xi v^2}(\partial_0 h)^2 - \frac{2M}{\xi Ze v}\nabla \cdot \Ab \partial_0h
- \frac{v^2}{2M}(\nabla \phi)^2\nn \\
&-\frac{Z^2e^2 v^2}{2M}\Ab^2
 -\frac{1}{2\xi}(\nabla \cdot \Ab)^2 -\frac{\xi Z^2e^2 v^4}{2M^2}\phi^2
-\frac{1}{4}F_{\mu\nu}F^{\mu\nu}+ \frac{1}{2}A^\mu  \Pi_{\mu\nu}A^\nu ,
\end{align} 
where we are not showing the rest of the Lagrangian, $\mathcal{L}_{\mathrm{higher}}$, which includes terms with more than two powers of the fields. Charge neutrality implies that $(v+h)^2 = n$.
We now choose unitary gauge by sending $\xi\rightarrow\infty$.  The advantage of unitary gauge is that it leaves only the ``physical'' fields in $\mathcal{L}_{\mathrm{quad}}$. The main effect of the gauge fixing with this choice of $\xi$ is to drop $\phi$ from the Lagrangian, as the term $\sim \xi \phi^2$ essentially gives it an infinite gap, causing it to decouple from the other fields.  At the same time, going to unitary gauge eliminates terms involving $\partial_{0}h$ from the Lagrangian.  To proceed, we also separate $A_\mu$ into three parts: $A_0$, $\Ab^\perp$ and $\Ab^L$, with $\Ab(\mathbf{p}) = \Ab^{\perp}(\mathbf{p} )+ \Ab^{L}(\mathbf{p})$, such that $\Ab^\perp(\mathbf{p}) \cdot \mathbf{p}=0$ and $\Ab^{\perp}(\mathbf{p}) \cdot \Ab^{L}(\mathbf{p}) = 0$. The quadratic part of the Lagrangian in unitary gauge is then given by
\bea
\mathcal{L}_{quad}  &=&2\mu v h +Ze(v^2-n)A_0 +
h\left[   \mu+\frac{\nabla^2}{2M}\right]h + \frac{1}{2} \Ab^\perp \left[
-\partial_0^2+\nabla^2-m_A^2+\Pi^T
\right]\Ab^\perp 
+ \frac{1}{2} A_0\left[
-\nabla^2+\Pi
\right]A_0\nn\\
&&+ \frac{1}{2}   A^L_i\left[
\delta_{ij}(-\partial_0^2+\nabla^2-m_A^2) -\partial_i \partial_j \left(1-\frac{\partial_0^2}{\nabla^4}\Pi\right)
\right]A^L_j
+
A_0\left[   \partial_0\partial_i\left(-1+\frac{\Pi}{\nabla^2}\right)\right]A^L_i
+2 Z e v h A_0,
\eea 
with $m_A^2=Z^{2} e^{2} v^{2}/M = Z^{2} 4\pi \alpha v^{2}/M$. 
The chemical potential $\mu$ and the expectation value of $\langle \psi \rangle=v$ must be chosen to guarantee charge neutrality which, at zero temperature, implies that $v^2=n$. This requires $\mu=0$ at tree level. The physical interpretation of this choice is transparent: there is no Coulomb contribution to the energy from one photon exchange, since the ion and electron charge density cancel each other. The Coulomb energy arises from fluctuations of charge densities, described by two or more photon exchanges. These effects first arise only at one-loop order and are small. In \cite{Bedaque:2010ph} the one-loop value of the chemical potential was estimated to be of order $\mu\sim \alpha/l$, but we will not use this result here.
 $\Pi^T$ only contributes  to the propagator of the transverse photon $\Ab_\perp$ and it describes magnetic effects due to the electrons, such as Landau damping. Such effects do not play an important role here so we will simply neglect them.
At quadratic order the two transverse modes decouple from $h, \Ab^{L}$ and $A_{0}$, and have the dispersion relation (assuming $\Pi^T$ is small)
\beq
p_0^2 = \mathbf{p}^2+m_A^2.
\eeq 
The fields $A_0$, $\Ab^L$ and $h$ mix so, in order to extract the spectrum of quasiparticles, we will integrate out two of them. The choice of which fields to eliminate is arbitrary. We integrate out $\Ab^L$ first. The propagator of $\Ab^L$ can be found through the easily verified formula
\beq
(A \delta_{ij}+B \partial_i \partial_j)^{-1} = \frac{1}{A}\left( \delta_{ij}-\frac{B}{A+B\nabla^2} \partial_i \partial_j \right),
\eeq  
and find
\beq
\mathcal{L}_{quad}  =
h\left[   \mu+\frac{\nabla^2}{2M}\right]h + \frac{1}{2} \Ab^\perp \left[
-\partial_0^2+\nabla^2-m_A^2+\Pi^T
\right]\Ab^\perp 
+ \frac{1}{2} A_0\left[m_A^2
\frac{\nabla^2-\Pi}{-\partial_0^2-m_A^2+\Pi\frac{\partial_0^2}{\nabla^2}}
\right]A_0
+2 Z e v h A_0.
\eeq 
Notice that if $h$ and $A_{0}$ had not coupled to $\Ab^{L}$, there would have been no time derivative in the $h, A_0, \Ab^L$ sector and no quasiparticle could arise from this sector.  However, $\Ab^{L}$ couples to $h$ and $A_0$ through the time-space component $\Pi_{0i}$ of the polarization tensor, which is required to be non-zero by gauge invariance if $\Pi_{00}$ is non-zero.
We can now integrate out either $A_0$ or $h$. Eliminating $A_0$ we find
\beq\label{eq:GL-quad}
\mathcal{L}_{quad}  =h \left[   
  \mu+\frac{\nabla^2}{2M}+2M\frac{-\partial_0^2-m_A^2+\Pi \partial_0^2/\nabla^2}{-\nabla^2+\Pi}
\right]h
 + \frac{1}{2} \Ab^\perp \left[
-\partial_0^2+\nabla^2-m_A^2
\right]\Ab^\perp.
\eeq 

The propagator $G_{h}(p)$ for $h$ is 
\begin{align}\label{eq:Gh}
G_{h}(p_{0},p) = \frac{p^{2}/2M}{p_0^2 - \left( \frac{\mathbf{p}^2}{2M}\right)^2 - \frac{\mathbf{p}^2 m_A^2}{\mathbf{p}^2+\Pi(p,p_{0})}}.
\end{align}
The dispersion relations for the modes excited by $h$ are given by the locations of the poles of $G_{h}(p_{0},p)$.  The fact that the residues at the poles have non-trivial momentum dependence has important consequences and will be discussed in Section~\ref{sec:Corrections}.  It turns out that $G_{h}$ has two poles describing (nearly) stable quasiparticles.  Setting $\mu$ to its tree-level value $\mu=0$, the positions of these poles are implicitly defined by
\beq\label{eq:dispersion}
p_0^2 = \left( \frac{\mathbf{p}^2}{2M}\right)^2 + \frac{\mathbf{p}^2 m_A^2}{\mathbf{p}^2+\Pi(p_{0},p)}.
\eeq If the electrons were not dynamical, we would have $\Pi=0$ and the mode would become gapped with $p_{0}(p=0) = m_{A}$.  This is a well known result \cite{Foldy:1961kx,Brueckner:1967vn,PhysRev.156.204} of the jellium model, which describes charged scalars interacting through an unscreened Coulomb potential in the presence of a static charge background to render the system electrically neutral. 

The combination of the generation of a gap for the transverse photons along with a massive scalar mode in the model when electrons are non-dynamical can be viewed is an example of the Anderson-Higgs mechanism. The electron sea, however, does contribute to the polarization tensor and alters the jellium model result. Already at the smallest density where the condensate is likely to appear $a_0/l\approx 12$, the Fermi energy is about $0.5\,\mathrm{MeV} \sim 6\times 10^9\ K$ but the temperature is below $80\,\mathrm{eV} \approx 10^6\ K$. Thus,  the relevant temperatures are far below the Fermi energy, and we can use the zero temperature polarization tensor. In general, $\Pi(p_0,\mathbf{p})$ has a complicated non-analytic dependence on $p_0$ and $\mathbf{p}$. The behavior of $\Pi(p_0,\mathbf{p})$ as $p_0$ and $\pb$ become small, the relevant regime for the low lying excitations we are after, depends on how the limit is approached.  The asymptotic forms of $\Pi$ that we will need can be found in e.g. \cite{Kalashnikov:1979cy,Altherr:1992mf,Manuel:1995td}
\begin{align}
&\lim_{\pb \to 0}\mathrm{Re}\Pi(\pb,p_0=0) = \lim_{\pb \to 0} \frac{2\alpha \mu_{e} k_{F}}{\pi}\left[ 1-\frac{k_{F}-\pb^2/4k_{F}}{p}\ln\left| \frac{1-p/2k_{F}}{1+p/2 k_{F}}\right|\right] 
=  \frac{4\alpha \mu_{e} k_{F}}{\pi},\nn\\
&\lim_{\pb \to 0}\mathrm{Im}\Pi(\pb, p_0=0) =0,\nn\\
&\lim_{p_{0} \to 0} \mathrm{Re}\Pi(\pb,p_0) \to -\frac{4\alpha k_{F}^3}{3\pi \mu_e}\frac{\pb^2}{p_0^2} ,\nn\\
&\lim_{p_{0} \to 0} \mathrm{Im}\Pi(\pb,p_0) =0,\nn\\
&\lim_{\pb \to 0} \mathrm{Re}\Pi(\pb,p_0=x v_{F} p) = \frac{2\alpha \mu_{e} k_{F}}{\pi} \left[  2-x \ln\left| \frac{1+x}{1-x} \right|\right],\nn\\
&\lim_{\pb \to 0} \mathrm{Im}\Pi(\pb,p_0=xv_F p)=2\alpha \mu_{e}k_{F} x\ \theta(1-x),
\end{align}  
where $\mu_e=\sqrt{k_F^2+m^2}$ is the electron chemical potential, $v_F=k_F/\mu_e$ is the Fermi velocity. The non-relativistic result valid at smaller densities is obtained from the above by setting $\mu_e\rightarrow m$ and $v_F\rightarrow k_F/m$.

By definition gapped quasiparticles have a finite energy $p_{0}$ as their momentum approaches zero. 
In this regime $\Pi$ is given by
\beq
\Pi \approx -\frac{4\pi\alpha Z n}{\mu_e}\frac{\mathbf{p}^2}{p_0^2}.
\eeq In the $\mathbf{p}\rightarrow 0$ limit, eq.~(\ref{eq:dispersion}) gives a pole at
\beq
p_0^2 = m_A^2 + \frac{e^2 Zn}{\mu_{e}} = 4\pi Z\alpha n\left(  \frac{Z}{M}+\frac{1}{\mu_{e}} \right) = \frac{4\pi Z\alpha n}{m_{red}},
\eeq  with $m_{red} = \mu_e  M/(Z\mu_e+ M)$ and $Z n=n_e$ is the electron density. This mode is gapped, justifying the assumption that $p_0$ remains finite as  $\mathbf{p}\rightarrow 0$. Its frequency is changed from the electron plasma frequency $\omega_{p}^{2}=4\pi\alpha n_e/m$ by (i) relativistic effects changing $m$ into $\mu_{e}$ and (ii) the change of the electron mass to the reduced mass, since the electrons are oscillating against the nuclei and not against an static positive background. Numerically we find
\bea
p_0 &\approx& 500\ \mathrm{eV} \left(1+\frac{m^2}{k_F^2}\right)^{-1/4}\frac{a_0}{l}\nn\\
       &\approx& 5.8\times 10^6\ \mathrm{K}\  \left(1+\frac{m^2}{k_F^2}\right)^{-1/4} \frac{a_0}{l}.
\eea

We can also look for gapless modes by assuming that $p_0 = x v_F p$ ($x=$constant)  as $\mathbf{p}\rightarrow 0$.  We have
\bea\label{eq:Pix}
\mathrm{Re} \Pi &=& m_s^2 \left(1-\frac{x}{2}\log\left|\frac{1+x}{1-x}\right| \right)=m_s^2 f(x),\nn\\
\mathrm{Im} \Pi &=& \frac{\pi}{2} m_s^2 x\Theta(1-x)
\eea 
with $m_s^2=4\alpha \mu_ek_F/\pi$. As $\mathbf{p}\rightarrow 0$ the $p^4$ term in eq.(\ref{eq:dispersion}) can be neglected, and we find
\beq
\frac{p_0^2}{p^2}=x^2 v_F^2 = \frac{m_A^2}{m_s^2 f(x)}
\eeq or
\beq
\label{eq:pole}
x^2 \left(1-\frac{x}{2}\log\left|\frac{1+x}{1-x}\right| \right) = \frac{m_A^2}{v_F^2 m_s^2}.
\eeq 
The left-hand side is positive for $x<0.8355$ and reaches a maximum of $\approx 0.211$ at $x=0.623$.  The association of the solution of eq.~\eqref{eq:pole} with a quasiparticle implicitly assumes that the imaginary part of the quasi-particle's energy (that is, the attenuation constant) is small, which as we will see would not be the case for large $x$.  Numerically, the parameter $\frac{m_A}{v_F m_s}$ is given by
\bea
\frac{m_A}{v_F m_s} &=& 0.8\ Z^{2/3}(1+\frac{m^2}{k_F^2})^{1/4}\alpha^{1/2}\sqrt{\frac{m}{M}}\sqrt{\frac{a_0}{l}}\nn\\
&\approx& 0.0012 (  1+\frac{m^2}{k_F^2})^{1/4}\sqrt{\frac{a_0}{l}},
\eea 
and, to a very good approximation $x\approx m_A/(v_F m_s)$ and is very small.  So eq.~\eqref{eq:pole} can be solved. The solution of eq.~(\ref{eq:dispersion}) with $\Pi$ given by  eq.~(\ref{eq:Pix}) can be found using an expansion in powers of the small parameter $x$, and is given by
 \begin{align}
 p_0=\frac{m_A}{m_s}\ p  -i \pi \frac{m_A^2}{4v_F m_s^2} p,
 \end{align}
The imaginary part is suppressed compared to the real part by a factor of $\pi x/(4v_{f})\ll1$, and this gapless mode is almost stable.  If we had used the Thomas-Fermi approximation for the screening of the electrons, $\Pi=m_{s}^{2}$, the imaginary part of the dispersion relation would not have been visible.  The imaginary part encodes the effects of inelastic scattering of this mode with particle-hole excitations of the electrons.   The size of the imaginary part is actually enhanced by the high density parameter $a_0/l$, but this enhancement is overwhelmed by the suppression coming from the mismatch between the rest masses of the ion and the electron.  It seems reasonable to think that the physical reason for the near-stability of the gapless mode is that collisions between it and the particle-hole excitations are mostly elastic thanks to the large mismatch between the masses of the helium nuclei and electrons.  
 
This gapless mode was not noticed in previous analyses of the system. The inclusion of the off-diagonal elements proportional to $\Pi_{00}$ in the polarization tensor, demanded by gauge invariance, are crucial for their existence. One interpretation of this mode may be as some kind of electron zero-sound-like excitation mixed with nuclear condensate excitations, but its velocity and attenuation rate behave differently than those of zero sound in Fermi liquids.  As we have already noted this mode would not survive if the electrons were not dynamical.  It would also not survive if the nuclei were not dynamical and the condensate were to vanish, since as is well-known an electron gas in a neutralizing positive non-dynamical background does not have a zero-sound-like mode - instead there are only the (gapped) plasma oscillations.  In section~\ref{sec:Corrections}, where we discuss the possible corrections to the dispersion relations, we will have more to say on the identity of this gapless mode.
  
The linear dispersion relation is valid for all relevant momenta at the temperatures we are interested in. In fact, the (real part of the) polarization $\Pi$, including the next order in the momentum expansion, has the form
 \beq
 \Pi(p_0=x v_F p, \pb) = m_s^2 \left[  f(x) + \frac{p^2}{k_F^2} g(x)+\cdots\right],
 \eeq 
 where $g(x)$ is a function of order one, assuming $\Pi(p_0=x v_F p, \pb)$ is analytic in $p^2$ around $p^2=0$.   Then the form of $\Pi$ we show above follows essentially from dimensional analysis, since the only relevant momentum scale in the loops defining $\Pi$ is $k_{F}$. The solution to eq.~(\ref{eq:dispersion}) is
 \beq
 p_0^2 = p^2 \frac{m_A^2}{m_s^2} - p^4 \left[ \frac{m_A^2}{m_s^4 f^2(x)\left(  1+\frac{m_s^2 g(x)}{k_F^2} \right) + \frac{1}{4M^2}}+\cdots  \right].
 \eeq The $p^4$ term becomes as important as the $p^2$ term for $p=\bar p\approx m_s^2 f(x)\approx m_s^2$. The energy of the gapless mode at this momentum is $\bar p_0 = m_A/m_s \bar p\approx m_A$. The scale $m_A$, however, is significantly higher than the temperatures where the condensate can exist. Numerically it is given by
 \bea
 m_A &=& \sqrt{3} Z \alpha^2 m \sqrt{\frac{m}{M}} \left(\frac{a_0}{l}\right)^{3/2}\nn\\
 &\approx&1.3\times 10^4\ K \left(\frac{a_0}{l}\right)^{3/2}
 \eea
which translates to a temperature of $2.6\times10^{6} K$ for an He WD with a central mass density of $5\times10^{5} g/cm^{3}$, or $a_{0}/l \sim 35$, which we will take as our reference density parameter for white dwarfs.

\section{Higher order corrections}
 \label{sec:Corrections}
Up to now our analysis was restricted to leading/tree level order.   We will now discuss higher order corrections in order to verify that our leading results are indeed reliable.  This is specially important for the gapless mode as no apparent symmetry exists that would protect this mode from picking up a gap when corrections are calculated.    As was the case in the context of the deuteron condensate discussed in \cite{Bedaque:2010ph}, we expect that the perturbative corrections to our results above will be in powers of $a_{0}/l$.  However, the existence of a gap for the gapless mode we found at tree level, even if small, could qualitatively change our conclusions.
 
 One way of organizing the perturbative expansion for our model is to integrate out all but the $h$ and $\Ab^T$ fields, as it was done in eq.~(\ref{eq:GL-quad}), but keep terms with higher powers of $h$. The residue of the propagator for the $h$ field, $G_h(p)$ in eq.~(\ref{eq:Gh}) has a non-trivial momentum structure. For this reason it is more convenient to work with the canonically-normalized field $H$ defined by $H = (-\nabla^{2}/M)^{-1/2} h$ whose propagator is 
 \begin{align}
G_{H}(p_{0},p) = \frac{1}{p_0^2 - \left( \frac{\mathbf{p}^2}{2M}\right)^2 - \frac{\mathbf{p}^2 m_A^2}{\mathbf{p}^2+\Pi(p,p_{0})}}
\approx
\frac{1}{p_0^2-\frac{m_A^2}{m_s^2}p^2 + \cdots},
\end{align} 
where the approximation above is valid near the propagator's pole for small momenta. The crucial point is to observe that the self-interactions of $H$ described by terms proportional to $H^3, H^4, \cdots$ always include derivatives. As a result, the action for the $H$ field has the symmetry $H\rightarrow H+\eta$, with $\eta$ a spacetime constant. This symmetry guarantees that a gap will not be generated at any order in perturbation theory (or even non-perturbatively). In fact, this is the same symmetry that guarantees that Nambu-Goldstone bosons remain gapless to all orders in the loop expansion. For all practical matters, $H$ seems to be a Nambu-Goldstone boson, despite the fact that the symmetry being spontaneously broken, the electromagnetic $U(1)$, is a local symmetry. Before commenting on this somewhat surprising result, let us verify that $H$ is indeed derivatively coupled.

\begin{figure}[tbp]
\centerline{ {\epsfxsize=3.0in\epsfbox{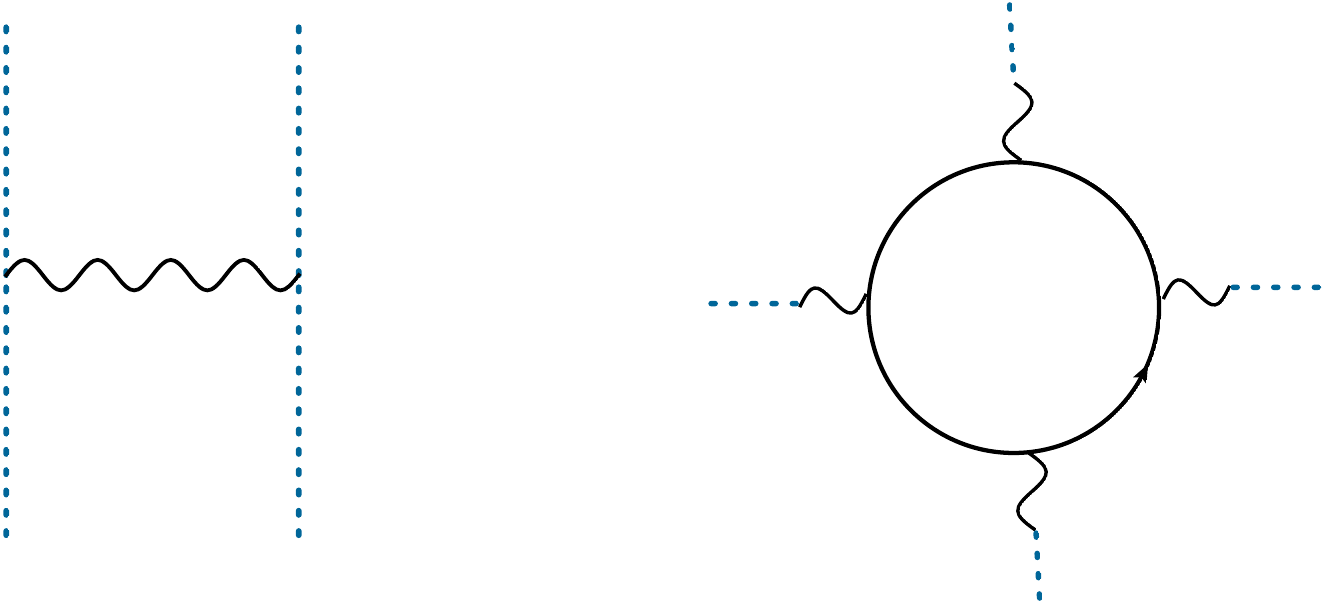}} }
 \noindent
\caption{(Color online.) Some of the graphs contributing to the self-interactions of the $h$ mode. Dotted lines represent the $h$ propagators, wavy lines are $A_0$ photons, while solid lines are electrons. }
\label{fig:h4}
\end{figure}  

We claim that a term involving a number $n$ of  $H$ fields contains at least $n$ derivatives, that is, they are schematically of the form $\partial^m H^n$, with $m>n$, where $\partial$ stands for a time or space derivative. If this statement were false, there would be terms in the effective action of $h$ of the form $\partial^{m-n} H^n$ with $m>n$, and we now argue that such terms are impossible. In fact, self-interaction terms for $h$ can be separated in two groups. One consists of terms generated by integrating out the fields $A_0$ and $\Ab^L$. An example is shown in Fig.~(\ref{fig:h4}). These diagrams, at small external momenta, generate terms proportional to $\alpha/m_s^2 \partial^{4} H^4$. Notice that, if the $A_0$ photons were not screened, diagrams like those in the left panel of Fig.~(\ref{fig:h4}) would generate terms exactly of the form we claim is impossible. The screening of $A_0$ photons is essential to this result. Another class of contributions is encoded in diagrams arising from the terms of order $A^3$ and higher in eq.~(\ref{eq:Selectron}), which describe interactions mediated by the polarization of the electron sea. An example is shown in the right panel of Fig.~(\ref{fig:h4}). Again, if the $A_0$ photons were massless, these diagrams would generate terms of the kind we claim is forbidden, but the screening of the Coulomb force eliminates this possibility.  In addition, it is conceivable that the non-localities of the fermion loop might generate negative powers of the external momenta. It is easy to see that this is not the case. In fact, a fermion loop with $n$ photon legs attached to them is proportional to~\footnote{Incidentally, this result in the $n=2$ case reproduces the value of $m_s^2\sim m k_F$.}
\bea
\sim \int d^4k \left(   \frac{1}{k_0+\mu_e-\frac{k^2}{2m}+i0k_0}\right)^n
\sim \frac{d^{n-1}}{d\mu_e^{n-1}}  \int d^4k  \frac{1}{k_0+\mu_e-\frac{k^2}{2m}+i0k_0}
\sim
 \frac{d^{n-1}}{d\mu_e^{n-1}} \underbrace{Zn}_{\stackrel{electron}{ density}}
 \sim
 m^{3/2} \mu_e^{3/2-n+1}
\eea (for simplicity we assumed non-relativistic electrons in showing this expression). The essential point is that the loops are infrared finite and the diagram does not diverge as the external momenta vanishes\footnote{The value of these loops depend, at small momenta, on the ratio $p_0/p$. But for all values of $p_0/p$ the zero-momentum limit is finite.}.
 
The conclusion we derive from this analysis is that, for all practical effects, the gapless mode we found is indistinguishable from a Nambu-Goldstone mode. This may seem at odds with the fact that the only symmetry which is spontaneously broken, the $U(1)$ gauge invariance, is actually a local symmetry. In relativistically invariant theories one would expect the Anderson-Higgs mechanism to play its usual role, and no physical gapless particle to arise if the symmetry being broken is local, but the situation is more subtle in non-relativistic theories.  In the non-relativistic case, it turns out that much rests on whether the electric forces mediated by the photons are screened.   Recall that  the Anderson-Higgs mechanism avoids the conclusion of the Goldstone theorem due to infrared divergences resulting from long-range forces mediated by unscreened photons. In our model, the screening due to the electrons destroys the long range forces, and the Goldstone theorem remains in effect. This phenomenon is actually well known, for a general discussion see ref.~\cite{Lange:1965zz,Guralnik:1964eu,Lange:1966zz}.  In a model where there is screening but no dynamical electrons, there would be a gapless mode, stable at tree level, as can be easily seen from eq.~(\ref{eq:dispersion}) with $\Pi=m_s^2$.  In our model the gapless mode can transfer energy to the electrons and decay already at tree level. Due to the difference in masses between ions and electrons this transfer is inefficient and the mode is almost stable.
 
 \section{Specific heat}
 \label{sec:SpecificHeat}
 
 As pointed out in Refs.~\cite{Gabadadze:2008mx,Gabadadze:2009jb,Gabadadze:2009dz} the main observable impact of having a nuclear condensate instead of an ion lattice or an ion gas inside white dwarfs is a change in the cooling curves caused by a change in the specific heat. The specific heat from the condensed phase at low temperatures comes both from the electrons  and the new gapless mode.  The contribution of the plasmon mode and the massive transverse photons is Boltzmann-suppressed. The electron contribution to the specific heat {\it per ion} originates from the first term in eq.~(\ref{eq:Selectron}) and is well known:
 \bea
 c_v^e &=&Z^{1/3} \left( \frac{\pi}{3}\right)^{2/3} \frac{\mu_e}{ n^{2/3}}T\nn\\
            &\approx& 0.19    \sqrt{1+\frac{m^2}{k_F^2}}  \frac{l}{a_0} \left( \frac{T}{10^6 \ K}\right).
 \eea 
 The gapless zero-sound-like mode contribution can be easily calculated through
 \beq
 c_v = \frac{1}{n}\frac{d\epsilon}{dT} =  \frac{1}{n}\frac{d}{dT}\int \frac{d^3p}{(2\pi)^3}\frac{c_{H}p}{e^{cp/T}-1}
 =\frac{1}{n} \frac{2\pi^2}{15 c_{H}^3}T^3
 \eeq for a mode with dispersion relation $p_0 = c_{H} p$. Numerically we have
 \bea
 c_v &=& \frac{16 \pi^{5/2}}{15\sqrt{3}}\frac{\alpha^{3/2}}{Z^2}\left( \frac{M}{m} \right)^{3/2}  \left(  \frac{l}{a_0} \right)^{9/2} \left(\frac{T}{\alpha^2 m}  \right)^3 \nn\\
 &\approx& 3.3\times 10^{4} (1+\frac{m^2}{k_F^2})^{3/4}  \left(  \frac{l}{a_0} \right)^{9/2}  \left( \frac{T}{10^6\ K}\right)^3.
 \eea
 
\begin{figure}[tbp]
\centerline{ {\epsfxsize=3.0in\epsfbox{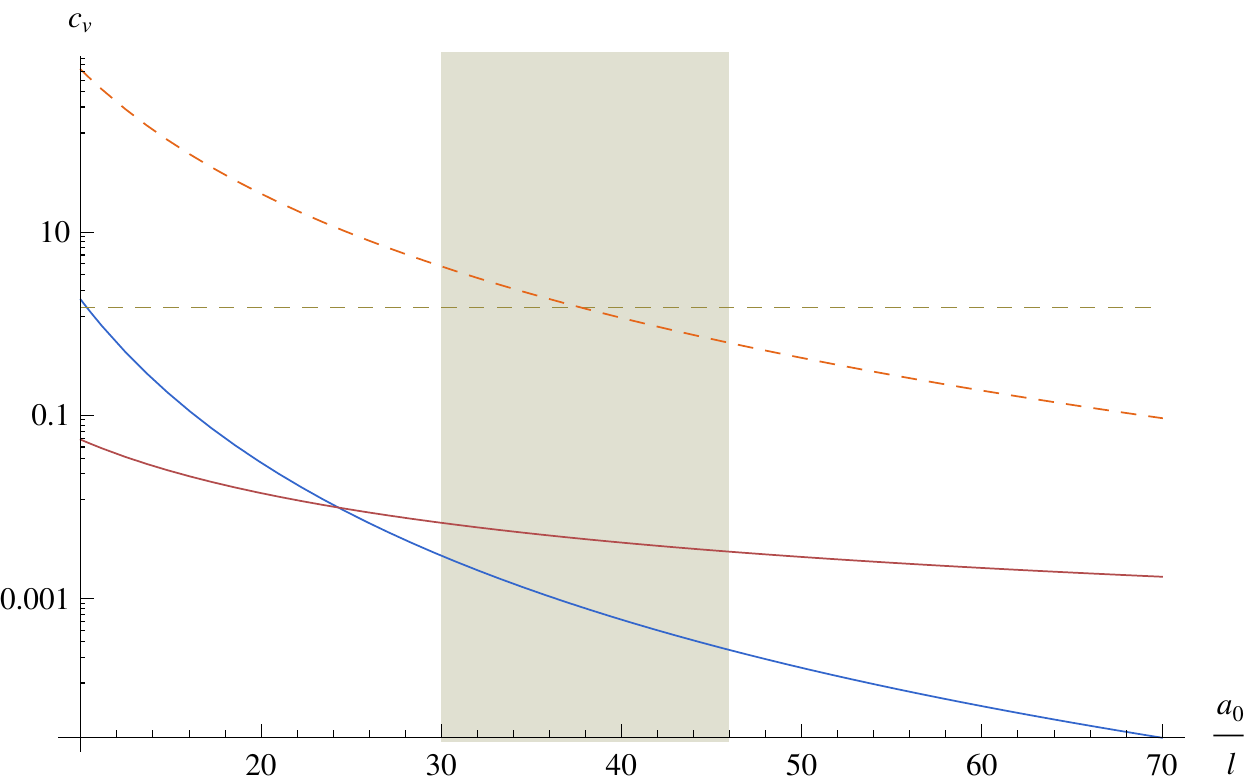}} }
 \noindent
\caption{(Color online.) Specific heat per ion from electrons (red) and nuclei (blue), with the temperature is held fixed at $T=5\times10^5$ K.   The band indicates the regions with densities relevant for He WDs, which range between $3\times 10^5\ g/cm^3$ and $ 10^6\ g/cm^3$.  The corresponding $T_{c}$ values range from $5.2\times10^{5} K$ to $\sim1.2\times10^{6} K$.  Also shown is the value for free ions ($3/2$ in yellow) and lattice phonons that would exist if the nuclei crystallized (dashed lines).  }
\label{fig:specificheat}
\end{figure}

The contribution of the gapless NG-like mode to the specific heat obviously has a very different temperature and density dependence than the contribution of the electrons, and which one is dominant depends on the specific parameters of interest.  In Fig.~(\ref{fig:specificheat}) we show the values of the electron and ion to the specific heats for the fixed temperature $T=5\times10^5$~K, at densities relevant for white dwarfs.  For this particular choice of parameters, the electron contribution happens to be dominant, but at higher temperatures the situation can be different. If instead of an nuclear condensate we had an ion lattice, as might be the case at low enough temperatures, the specific heat would be dominated by the lattice phonon contribution. This contribution has been analyzed before\cite{shapiro1983black} and is given by
\bea
c_v^{lattice} &=&\frac{16\pi^4}{5}  \left(\frac{T}{\theta_D}\right)^3\nn\\
&=&  \frac{16 \pi^{4}}{15\sqrt{3}}\frac{1}{Z^3} \left( \frac{M}{m} \right)^{3/2}  \left(  \frac{l}{a_0} \right)^{9/2} \left(\frac{T}{\alpha^2 m}  \right)^3   \nn\\
&\approx&
1.5\times 10^{8} \left(  \frac{l}{a_0} \right)^{9/2} \left(\frac{T}{10^6\ K}\right)^3,
\eea where the Debye temperature $\theta_D$ is the same as what we called $m_A$, and the formula given above is valid for $T\ll \theta_{D}$. This contribution is  enhanced, as compared to the condensed phase, by a power of $1/\alpha^{3/2}$. The specific heats from the different sources are plotted in Fig.~(\ref{fig:specificheat}).  The picture for the specific heat of the nuclear condensate we have found broadly agrees with the results of \cite{Gabadadze:2008mx,Gabadadze:2009jb,Gabadadze:2009dz}.  Taking into account the contributions of all relevant quasiparticles, we see that if nuclear condensation takes place in the cores He WDs, their specific heat of the cores will be much lower than if the helium were in a crystalline or plasma phase.

\section{Conclusions}
 \label{sec:Conclusions}
 
The quasiparticle spectrum of the nuclear condensate turns out to be unexpectedly rich. In addition to transverse photons picking up a magnetic mass through the Higgs mechanism and the existence of a gapped plasmon quasiparticle, we also found that the system supports an almost stable gapless mode.  Somewhat surprisingly, despite the fact that the symmetry being broken is a local one, this physical gapless mode can be identified as a Nambu-Goldstone mode, arising because in our model the condensing field is non-relativistic and because the electrons screen the long-range Coulomb forces.   To find this mode it turned out to be crucial to take into account the constraints of gauge invariance on the electron contribution to the photon polarization tensor.   This gapless excitation has a number of interesting properties from the perspective of field theory, and their further exploration is a promising direction for future work.  For phenomenology it is important to explicitly compute the finite-T corrections to the spectrum, since the relevant temperatures for He WDs are not so widely separated from $T_{c}$, and finite-T corrections could be important for some quantities.   An effect we did not consider in this paper is that at low enough temperatures the electrons will pair up and condense, and it would be nice to understand better when such phenomena become relevant.  It would also be interesting to check whether there is an analogous mode in super-dense deuterium\cite{Berezhiani:2010db,Bedaque:2010ph}, as seems likely, and to understand its phenomenological consequences.
 
With the spectrum of quasiparticle excitations in hand, we briefly examined the specific heat of the nuclear condensate, comparing our results to previous work in the literature.  For the parameter values appropriate for describing helium white dwarfs, the contribution of the gapless (Nambu-Goldstone-like) mode to the specific heat can be larger or smaller than the contributions from the particle-hole excitations of the degenerate electron gas, but generically both contributions are very small compared to that of an uncondensed plasma.  Our results suggest that the specific heat of the core of a white dwarf will drop by a factor of $\approx 10^{-2}$ upon ion condensation, similarly to previous analyses\cite{Gabadadze:2009dz}.

It remains an interesting open question whether such a change in the thermodynamic properties of the core of a white dwarf would have observable consequences.  This matter was previously discussed in Ref.~\cite{Gabadadze:2009dz}, which argued that the consequences would be detectable based on a simple model, and in Ref.~\cite{Benvenuto:2011fj}, which argued that observable effects were unlikely based on a somewhat more detailed model.  The ultimate answer to this question depends on developing a detailed model of white dwarf cooling in the presence of ion condensation, taking into account all relevant features of the system, such as {\it e.g.} the phenomenological consequences  of the suggestion of Ref.~\cite{Rosen:2010es} that the transition to the condensed phase is first order rather than second order, or the implications of the complete quasi-particle spectrum on e.g. the rate of neutrino emission\cite{2004ApJ...602L.109W}.  Aside from looking at the effects of condensation on the cooling rates of He WDs, it may also be interesting to consider whether the way electric and magnetic fields are screened in the nuclear condensate may have any observable consequences for He WDs\cite{Gabadadze:2009qe,Mirbabayi:2010ia}.

\acknowledgements{We thank Mike Buchoff, David B. Kaplan, Dean Lee, Sanjay Reddy, Thomas Schaefer and Dam Son for very helpful conversations, and thank Mike Buchoff and Srimoyee Sen for collaboration on related projects.  P.~F.~B. and E.~B. thank the US DOE for support under DOE grant DE-FG02-93ER-40726, and E.~B. also thanks Jefferson Science Associates for support under the JSA/JLab Graduate Fellowship program.  A.~C. is grateful to STFC for support through the HEP group grant at DAMTP.}	

 \section{Appendix: Polarization tensor in the absence of boost invariance }
 In the main text we used relations between the different components of the polarization tensor $\Pi_{\mu\nu}$ imposed by gauge symmetry even in the absence of boost invariance. In particular we used the fact that, in the absence of boost invariance, $\Pi_{\mu\nu}$ depends on two arbitrary functions of $p_0$ and $p^{2}=p_{i}p^{i}, i=1,2,3$. In order to see this, consider the most general
 symmetric tensor, function of $p_\mu$ and $n_\mu$ (where  $n_\mu=(1,0,0,0)$ points in the time direction):
 \begin{align}
 \Pi_{\mu\nu} = A(p_0, p^2)  p_\mu p_\nu + 
 B(p_0, p^2) n_\mu n_\nu
 +C(p_0, p^2) (p_\mu n_\nu+p_\nu n_\mu)
 +D(p_0, p^2) g_{\mu\nu},
 \end{align}
 with $p_0=p_\mu n^\mu$. The Ward identity implies in
 \begin{align}
0= p^\mu \Pi_{\mu\nu} = A(p_0, p^2)  p^2 p_{\nu} + 
 B(p_0, p^2) p_0 n_\nu
 +C(p_0, p^2) (p^2 n_\nu+p_\nu p_0)
 +D(p_0, p^2) p_\nu.
 \end{align}
 This means that we can eliminate two of the four arbitrary functions using the relations
 \bea
 A(p_0, p^2)  p^2 +C(p_0, p^2) p_0 +D(p_0, p^2)=0,\nn\\
  B(p_0, p^2) p_0+C(p_0, p^2) p^2=0.
 \eea


\bibliography{nuclear_liquids,astro_refs}
\end{document}